# The adjustment-stabilization method for constrained systems

Wen-biao Han [a,b,*], Xin-hao Liao [a]

[a] *Shanghai Astronomical Observatory, Chinese Academy of Sciences, Shanghai 200030, China*
[b] *Graduate School of the Chinese Academy of Sciences, Beijing 100049, China*

**Abstract**

For constrained system which has several independent first integrals, we give a new stabilization method which named *adjustment-stabilization method*. It can stabilize all known constants of motion for a given dynamical system very well instead of the stabilization and post-stabilization methods which only conserves one of all first integrals. Further more, new method can improve numerical accuracy too. We also point out the post-stabilization is just a simplest case of the new method.

## 1. Introduction

For a given dynamical system which has several independent first integrals, in this paper we also call constrained system, how to conserve these constants of motion in numerical computing is a very important problem. The traditional numerical algorithms single-step and multi-steps method can attain a good accuracy in not too long time, but they all introduce inherent numerical dissipation, so that the errors of the numerical algorithm yield a great bias from the variant as long as the integrated time increase. This serious problem will bring bad results when we try to study the long evolution or the characters of a physical model. For example, when the traditional methods are used to resolve the simple two-body problem, the numerical trajectory all move to inner or outer. In all these integrations it was found that the energy and angle momentum error grew linearly with time and, because the mean motions depend on the energy, the longitude error grew quadratically with time.

To find favorite numerical algorithms, there were two main aspects. One was first founded by Kang Feng et al. [1] called Symplectic integration algorithms (SIAs), they can be applied to Hamilton systems because every transform of the method is symplectic, so they preserve the symplectic structure of the Hamilton systems. It is the best method to integrate numerically for Hamilton system, it can conserve the energy of system very good. SIAs have been widely used and greatly developed [2–18]. However, SIAs have some weakness. Firstly, they are only limited to Hamilton systems. Especially, when we want gain explicit SIAs, the Hamilton functions must be separable like $H(p, q) = f(p) + g(q)$. Secondly, the adaptively adopted SIAs are low-order in general. In this sense, it necessary to develop a conventional high-order integration method with the conservation of general constrained systems. On the other side, somebody corrected the errors of the invariant integrations by add the control term to the right functions or at the end of every step of computing. For example, there is the *stabilization of differential equations* of Baumgarte [19] or the *manifold correction* of Nacozy [20]. As to the stabilization method [19], one main drawback lies in the question of the best choice of the stabilizing parameters. The *manifold correction* of Nacozy [20] can overcome the demerit and is of convenience in application. As an illustration, it is termed the *post-stabilization* by Chin [21] or the *projection method* by Hairer et al. [22], in this paper we call this kind method *post-stabilization*. For either Baumgarte's method or the *post-stabilization*, usually the effectiveness of stabilization becomes better by choosing an energy integral as the control term. In this case, they are also regarded to as *energy stabilization*. For the Kepler problem, the, *energy stabilization* can make the longitude error grew linear with time [23–26,28].

---

\* Corresponding author.
 *E-mail address:* wbhan@shao.ac.cn (W.-b. Han).

But the *post-stabilization* can only conserve one first integral (energy usually) for constrains systems which have several independent invariant integrals in general, this will be illustrated in this paper later. For overcome this disadvantage, we present a new method with the *least squares adjustment* theory. For this reason, we call the new method *adjustment-stabilization*. It considers all known independent conservations of the system to conditional equations, distributes the errors of invariants to every variables most reasonable. The *adjustment-stabilization* can preserve all the known integrals of motion and give better accuracy. If we only choose one constraint and all numerical algorithms of every variable are same (having equal accuracy), it will go back to the *post-stabilization*. So, in our views, the *post-stabilization* method is a special example of the *adjustment-stabilization*.

The paper is organized as follows. In Section 2 we deduce the mathematical formulas of the *adjustment-stabilization*. We do several numerical experiments to explore its effect compare to *post-stabilization* in Section 3. At last, we present our conclusion and discussion.

## 2. Mathematical formulas

For a $N$ dimensions dynamical system which has $m$ independent first integrals

$$\dot{x} = f(x, t), \tag{1}$$

where $x$ and $\dot{x}$ is $N$ dimensions column vector.

The $m$ conservations are

$$\phi_1(x) = c_1, \phi_2(x) = c_2, \ldots, \phi_m(x) = c_m. \tag{2}$$

Let $\Phi_j(x) = \phi_j - c_j$, $j = 1, 2, \ldots, m$, can be express $m$ dimensions column vector, and the above equation becomes

$$\Phi(x) = 0. \tag{3}$$

If at the time $t_n$, having the real solution $x_n$ of system, then

$$\Phi(x_n) = 0, \tag{4}$$

suppose having numerical solution $\tilde{x}_{n+1}$ at the next step $t_{n+1}$, in general

$$\Phi(\tilde{x}_{n+1}) \neq 0, \tag{5}$$

but a small quantity. The step size is $h = t_{n+1} - t_n$.

If we consider the numerical solution $\tilde{x}_{n+1}$ to a "measure value" of the real solution $x_{n+1}$ at the time $t_{n+1}$, and regard Eq. (3) as the conditional equation in the theory of adjustment. Then we can obtain the best modified value $\bar{x}_{n+1}$ of $\tilde{x}_{n+1}$, it is a better approximate to real solution, and makes

$$\Phi(\bar{x}_{n+1}) = \begin{cases} 0 & \Phi(x) \text{ is linear,} \\ \text{higher order} & \Phi(x) \text{ is nonlinear.} \end{cases} \tag{6}$$

Now we deduce the mathematical formulas of *adjustment-stabilization*.

Suppose at time $t_{n+1}$, the real solution and the numerical solution have relation as follows

$$x_{n+1} = \bar{x}_{n+1} + v_{n+1}, \tag{7}$$

where the $v_{n+1}$ is a $N$ dimensions column vector called corrected number. Applying Eq. (7) to (3), we have

$$A v_{n+1} + W_{n+1} = 0, \tag{8}$$

where $A$ is $m \times N$ dimensions matrix which has form

$$A = \begin{pmatrix} \frac{\partial \Phi_1}{\partial x_1} & \cdots & \frac{\partial \Phi_1}{\partial x_N} \\ \vdots & \ddots & \vdots \\ \frac{\partial \Phi_m}{\partial x_1} & \cdots & \frac{\partial \Phi_m}{\partial x_N} \end{pmatrix}. \tag{9}$$

And the $W_{n+1}$ is $m$ column vector called closure error

$$W_{n+1} = \Phi(\tilde{x}_{n+1}). \tag{10}$$

Suppose the accuracy-weight-matrix of $N$ numerical methods for Eq. (1) is $P$,

$$P^{-1} = Q = \frac{1}{\sigma_0^2} \begin{pmatrix} \sigma_1^2 & \sigma_{12} & \cdots & \sigma_{1N} \\ \sigma_{12} & \sigma_2^2 & \cdots & \sigma_{2N} \\ \cdots & \cdots & \cdots & \cdots \\ \sigma_{N1} & \sigma_{N2} & \cdots & \sigma_N^2 \end{pmatrix}. \tag{11}$$

Here, $\sigma_0$ is a unit mean error. The diagonal elements indicate variances of $N$ numerical methods for a same calculation, $\sigma_{ij}$ denotes degree of the correlation of two numerical algorithms, obviously, if we adopt $N$ independent methods, $P$ becomes a diagonal matrix, the element of matrix,

$$p_i = \frac{\sigma_0^2}{\sigma_i^2} \tag{12}$$

presents the $i$th method's accuracy comparing to others, the bigger $p_i$ states better accuracy. And now, we have,

$$P = \begin{pmatrix} p_1 & \cdots & 0 \\ \vdots & \ddots & \vdots \\ 0 & \cdots & p_N \end{pmatrix}. \tag{13}$$

If all the $N$ numerical methods are same and independently, then $P$ is $N \times N$ unit matrix, this is a simplest case but is the most important, and we suggest to use it. In this paper, we adopt the same method for $N$ components of Eq. (1) independently, so $P$ is a unit matrix.

The most complex is these methods are dependent each other, then the matrix is not diagonal. And we strongly advise not to adopt this instance, because there is very difficult to determine the values of the off-diagonal element of $P$ quantitatively. Someone perhaps chooses different algorithms for components of Eq. (1) but independently, so must do numerical test to determine $P$, this is also complicated and little useful.

According to the least squares theory, the best modification requires the best error distribution should make the quadratic sum of the departure of "measure value" to modified value smallest. So, we have,

$$v_{n+1}^{\mathrm{T}} P v_{n+1} = \min. \tag{14}$$

Associating Eqs. (8) and (14), we can find a conditional minimax problem, so applying the Lagrange multiplicator method, supposing the multiplicators are $\mathbf{K} = (k_1, k_2, \ldots, k_m)^{\mathrm{T}}$ called correlated number vector. We can construct the function

$$\mathbf{\Theta} = \mathbf{v}_{n+1}^{\mathrm{T}} \mathbf{P} \mathbf{v}_{n+1} - 2\mathbf{K}^{\mathrm{T}}(\mathbf{A}\mathbf{v}_{n+1} + \mathbf{W}_{n+1}). \tag{15}$$

For satisfying Eq. (12), requiring

$$\frac{\mathrm{d}\mathbf{\Theta}}{\mathrm{d}\mathbf{v}} = 2\mathbf{v}^{\mathrm{T}} \mathbf{P} - 2\mathbf{K}^{\mathrm{T}} \mathbf{A} = \mathbf{0}. \tag{16}$$

We sometimes do not write the subscript $n+1$ from here for convenience. From the above equation, we can solve the corrected number $\mathbf{v}$

$$\mathbf{v} = \mathbf{P}^{-1} \mathbf{A}^{\mathrm{T}} \mathbf{K}. \tag{17}$$

So we can decide the $N$ components of corrected number $\mathbf{v}$ by Eqs. (17) and (8)

$$\mathbf{v} = -\mathbf{P}^{-1} \mathbf{A}^{\mathrm{T}} (\mathbf{A}\mathbf{P}^{-1}\mathbf{A}^{\mathrm{T}})^{-1} \mathbf{\Phi}(\tilde{\mathbf{x}}_{n+1}). \tag{18}$$

Obviously the solution of $\mathbf{v}$ cannot only eliminates the closure error $\mathbf{W}$ but also make the distribution of error is optimized.

Now we can give the process of *adjustment-stabilization* as follows in detail.

(a) Suppose we have the real solution of system at $t_n$, then we obtain the numerical solution $\tilde{\mathbf{x}}_{n+1}$ at $t_{n+1}$ by certain numerical algorithms;
(b) Modify $\tilde{\mathbf{x}}_{n+1}$ by the formula

$$\bar{\mathbf{x}}_{n+1} = \tilde{\mathbf{x}}_{n+1} - \mathbf{P}^{-1}\mathbf{A}^{\mathrm{T}}(\mathbf{A}\mathbf{P}^{-1}\mathbf{A}^{\mathrm{T}})^{-1} \mathbf{\Phi}(\tilde{\mathbf{x}}_{n+1}); \tag{19}$$

(c) Regard the $\bar{\mathbf{x}}_{n+1}$ as the real solution at time $t_{n+1}$ and continue computing next step.

If we choose $N$ same numerical algorithms independently each other for $N$ components of $\mathbf{x}$, and only consider one conservation of system, then Eq. (17) becomes

$$\bar{\mathbf{x}}_{n+1} = \tilde{\mathbf{x}}_{n+1} - \mathbf{\Phi}(\tilde{\mathbf{x}}_{n+1}) \frac{\partial \mathbf{\Phi}}{\partial \mathbf{x}} \bigg/ \left( \frac{\partial \mathbf{\Phi}}{\partial \mathbf{x}} \bullet \frac{\partial \mathbf{\Phi}}{\partial \mathbf{x}} \right) \bigg|_{\mathbf{x}=\tilde{\mathbf{x}}_{n+1}}, \tag{20}$$

where the $\bullet$ represents the Euclidean inner product. This is just the post-stabilization method given by chin [21]. So we can view the *post-stabilization* is the most simple example of the *adjustment-stabilization*.

The disposal of *adjustment-stabilization* happens to maintain double the order of the integrals $\mathbf{\Phi}$ accuracy [21,27,28]. By Taylor's expansion around $\tilde{\mathbf{x}}_{n+1}$, one has

$$\begin{aligned}
&\phi_j(t_{n+1}, \bar{\mathbf{x}}_{n+1}) \\
&= \phi_j(t_{n+1}, \tilde{\mathbf{x}}_{n+1}) - \frac{\phi_j(t_{n+1}, \tilde{\mathbf{x}}_{n+1}) - c_j}{\nabla^2 \Phi_j(t_{n+1}, \tilde{\mathbf{x}}_{n+1})} \bullet \nabla \Phi_j(t_{n+1}, \tilde{\mathbf{x}}_{n+1}) \\
&\approx \phi_j(t_{n+1}, \tilde{\mathbf{x}}_{n+1}) - (\phi_j(t_{n+1}, \tilde{\mathbf{x}}_{n+1}) - c_j)^2 \\
&\quad + (\phi_j(t_{n+1}, \tilde{\mathbf{x}}_{n+1}) - c_j)^2 \frac{(\nabla \Phi_j)^{\mathrm{T}} \Phi_{j(\mathbf{xx})} \nabla \Phi_j}{2(\nabla^2 \Phi_j)^2} \bigg|_{\mathbf{x}=\tilde{\mathbf{x}}_{n+1}} \\
&= c_j + (\phi_j(t_{n+1}, \tilde{\mathbf{x}}_{n+1}) - c_j)^2 \frac{(\nabla \Phi_j)^{\mathrm{T}} \Phi_{j(\mathbf{xx})} \nabla \Phi_j}{2(\nabla^2 \Phi_j)^2} \bigg|_{\mathbf{x}=\tilde{\mathbf{x}}_{n+1}},
\end{aligned}$$

where $j = 1, 2, \ldots, m$, that is

$$\mathbf{\Phi}(\bar{\mathbf{x}}_{n+1}) \sim \mathbf{\Phi}^2(\tilde{\mathbf{x}}_{n+1}). \tag{21}$$

If the $\Phi_j$ is a linear function of $\mathbf{x}$, then $\Phi_{j(\mathbf{xx})}$ is zero, so has $\mathbf{\Phi}(\bar{\mathbf{x}}_{n+1}) = \mathbf{0}$, just preserves the conservations of system strictly.

In Section 3 we compare *adjustment-stabilization* with *post-stabilization* make use of two numerical experiments. The physics models we use are the two-body problem, and two-dimensional nonlinear oscillator.

## 3. Numerical examples

In this section, we compare the *adjustment-stabilization* with *post-stabilization*. We choose the all numerical methods for $N$ components of $\mathbf{x}$ are same and independently each other, then the $\mathbf{P}$ is $N \times N$ unit matrix.

### 3.1. The Kepler problem

For the simple Kepler problem

$$\begin{cases} \dot{q}_i = p_i, \\ \dot{p}_i = -\frac{q_i}{r^3}, \end{cases} \tag{22}$$

where $i = 1, 2, 3$, $r = \sqrt{q_1^2 + q_2^2 + q_3^2}$. Here we consider two first integrals

$$\begin{cases} H = \frac{1}{2} p^2 - 1/r, \\ h = |\vec{r} \times \dot{\vec{r}}| \end{cases} \tag{23}$$

are energy and angle momentum respectively, the *adjustment-stabilization* adopts the two integrals and the *post-stabilization* by the energy. The numerical integrator is Runge–Kutta78. In Figs. 1–3 we present the energy, angle momentum and longitude errors of three methods: the Runge–Kutta78 only, RK78 by *post-stabilization* and RK78 by *adjustment-stabilization*. The eccentricity of the orbit is 0.01, the step size is $T/50$, where $T$ is periodicity of motion. In Table 1, we compare three methods in different step size, and give the CPU time too. In Table 2, we adopt 6th order Adams–Bashforth (multi-step method) to replace RK78, and repeat process above.

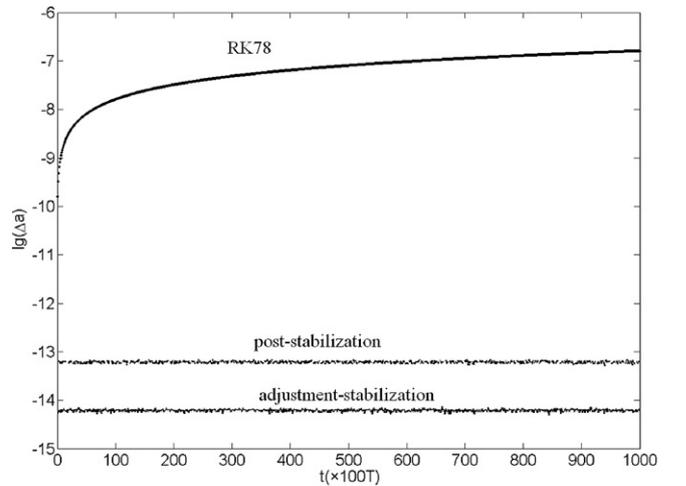

Fig. 1. The errors of semi-major axis. For post-stabilization, we replace $\Delta a$ by $10 \times \Delta a$. In fact, post-stabilization and adjust-stabilization are almost the same, and both the two methods are good at control of energy. This figure is about Kepler problem.

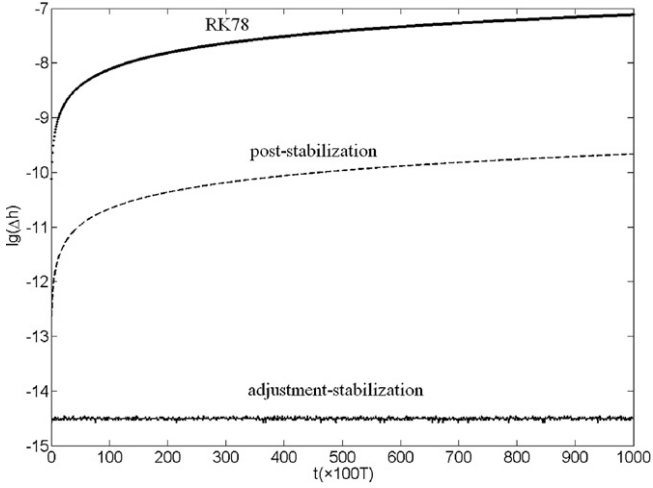

Fig. 2. The logarithm of angle momentum errors. We can find obviously the new method is much better. This figure is about Kepler problem.

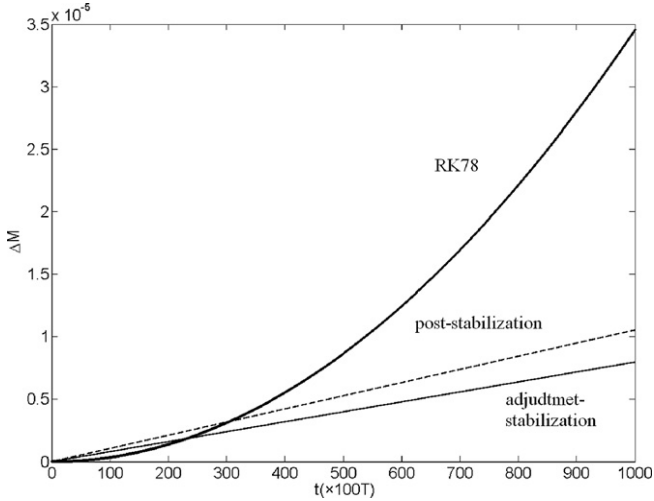

Fig. 3. For RK78, we replace $\Delta M$ by $\Delta M \times (5 \times 10^{-4})$. The longitude errors are both linear with time except RK78, but new method has better accuracy. This figure is about Kepler problem.

From these pictures and tables, the *adjustment-stabilization* by several conservations can preserve every known integral very good and improve the accuracy comparing to the *post-stabilization*, and do not cost too much CPU time.

### 3.2. Two-dimensional nonlinear oscillator

In the section, we apply classical Runge–Kutta4 and 4th order Adams–Bashforth integrators with the *adjustment-stabilization* and *post-stabilization* for compare.

For a nonlinear two-dimensional oscillator

$$H = \frac{1}{2}(p_1^2 + p_2^2 + q_1^2 + q_2^2) + q_1^2 q_2 + \frac{1}{3}q_2^3, \quad (24)$$

Table 1
For Kepler problem. The basic integrator is RK78, M1 is post-stabilization, and M2 is the new method. The integrate time is $10^5\ T$, Table 2 is same

| Step-size | Method | $|\Delta E/E|$ | $|\Delta h/h|$ | $|\Delta M/M|$ | CPU time (s) |
|---|---|---|---|---|---|
| $T/25$ | RK78 | 8.3E−5 | 4.0E−5 | o(1) | 22.3 |
| | M1 | 5.8E−15 | 1.1E−7 | 2.6E−3 | 23.4 |
| | M2 | 6.4E−15 | 3.3E−15 | 2.1E−3 | 24.3 |
| $T/50$ | RK78 | 1.6E−7 | 7.7E−8 | 6.9E−2 | 44.1 |
| | M1 | 6.7E−15 | 2.2E−10 | 1.1E−5 | 45.9 |
| | M2 | 6.0E−15 | 3.1E−15 | 8.0E−6 | 49.5 |
| $T/100$ | RK78 | 3.1E−10 | 1.5E−10 | 1.3E−4 | 88.6 |
| | M1 | 6.0E−15 | 4.3E−13 | 3.6E−8 | 92.0 |
| | M2 | 6.2E−15 | 3.3E−15 | 2.5E−8 | 95.9 |

Table 2
For Kepler problem. The basic integrator is 6th order Adams–Bashforth (labeled as Adams)

| Step-size | Method | $|\Delta E/E|$ | $|\Delta h/h|$ | $|\Delta M/M|$ | CPU time (s) |
|---|---|---|---|---|---|
| $T/300$ | Adams | 4.0E−6 | 1.9E−6 | o(1) | 27.5 |
| | M1 | 6.4E−15 | 4.5E−9 | 3.6E−4 | 41.5 |
| | M2 | 5.1E−15 | 2.9E−15 | 1.3E−4 | 50.4 |
| $T/400$ | Adams | 5.4E−7 | 2.5E−7 | 2.3E−1 | 36.9 |
| | M1 | 6.4E−15 | 6.0E−10 | 6.4E−5 | 56.4 |
| | M2 | 6.7E−15 | 3.3E−15 | 2.3E−6 | 67.4 |
| $T/500$ | Adams | 1.1E−7 | 5.3E−8 | 4.8E−2 | 45.9 |
| | M1 | 6.2E−15 | 1.3E−10 | 1.7E−5 | 69.4 |
| | M2 | 5.6E−15 | 3.1E−15 | 6.0E−6 | 83.7 |

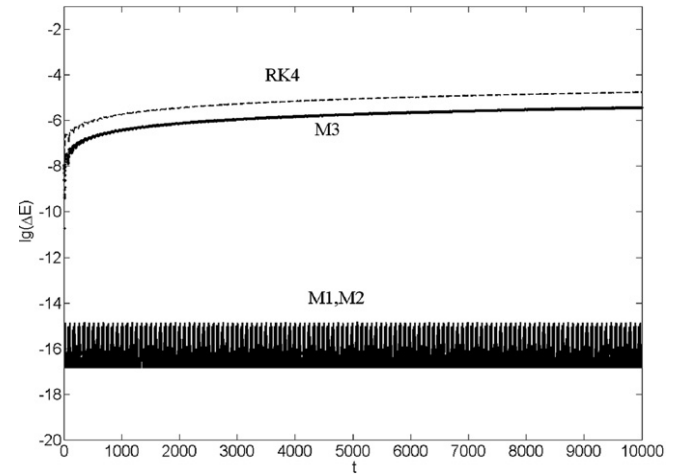

Fig. 4. Error of energy. M1 (adjustment-stabilization) and M2 (post-stabilization by energy) are almost the same, but M3 (post-stabilization by $F$) is not good. This figure is about non-harmonic oscillator.

there is another independent conservation

$$F = \frac{1}{2}(p_1 + p_2)^2 + \left(\frac{1}{2} + \frac{1}{3}(q_1 + q_2)\right)(q_1 + q_2)^2 \quad (25)$$

the *adjustment-stabilization* adopt control terms of (24) and (25), but the *post-stabilization* is just by the integral of (24) or (25). In Figs. 4 and 5, the M2, M3 present the *post-stabilization* by the energy $H$ and constant $F$ individual, M1 presents the RK4 with *adjustment-stabilization* by $H$ and $F$. We can find clearly new method preserves the two integrals and *post-stabilization* only preserves one. We put the initial conditions $q_1 = q_2 = p_1 = 0.1$, $p_2 = 0.4$, step size $h = 0.1$, and the total integration time $10^3$. Tables 3 and 4 give the numerical conclusion in different step size by RK4 and Adams–Bashforth correspondingly.

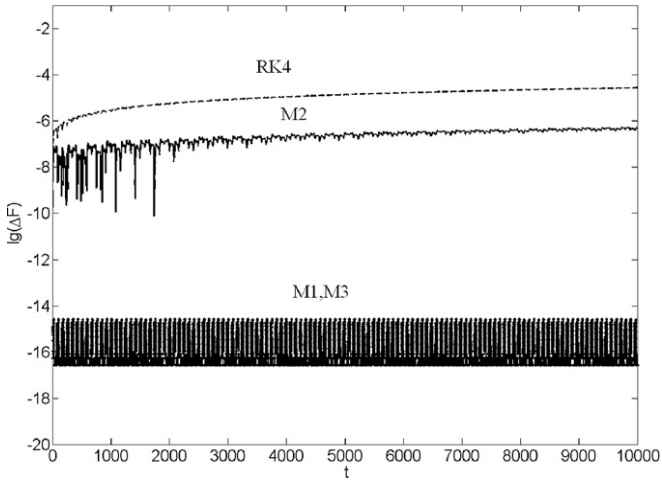

Fig. 5. Error of $F$. M1 (adjustment-stabilization) and M3 (post-stabilization by $F$) are almost the same, but M2 (post-stabilization by energy) is not good. This figure is about non-harmonic oscillator.

Table 3
For non-harmonic oscillator. The basic integrator is RK4. M1 presents adjustment-stabilization, M2 and M3 present post-stabilization by $E$ and $F$, respectively

| Step-size | Method | $|\Delta E/E|$ | $|\Delta F/F|$ | CPU time (s) |
|---|---|---|---|---|
| 0.1 | RK4 | 1.7E−5 | 2.7E−5 | 2.1 |
| | M2 | 6.9E−17 | 4.7E−7 | 2.1 |
| | M3 | 3.7E−6 | 2.8E−17 | 2.2 |
| | M1 | 9.7E−17 | 8.0E−17 | 2.3 |
| 0.05 | RK4 | 5.4E−7 | 8.5E−7 | 4.0 |
| | M2 | 2.8E−17 | 1.5E−8 | 4.0 |
| | M3 | 1.1E−7 | 0 | 4.1 |
| | M1 | 0 | 2.8E−17 | 4.1 |
| 0.01 | RK4 | 1.8E−10 | 2.9E−10 | 19.5 |
| | M2 | 1.4E−17 | 5.7E−12 | 19.5 |
| | M3 | 3.6E−11 | 2.8E−17 | 20.5 |
| | M1 | 4.1E−17 | 2.8E−17 | 20.7 |

## 4. Summary and discussion

For a dynamical system which has several independent first integral, the *post-stabilization* method just adopts one constant of motion as the stabilizing term, and only preserves one it in general. The *adjustment-stabilization* method chooses all known preservations as conditional equation, can maintain all of them very well and improve the accuracy. By numerical examples, we find the *adjustment-stabilization* method do not cost too much CPU time against the *post-stabilization*.

Table 4
For non-harmonic oscillator. The basic integrator is 4th order Adams–Bashforth. (labeled as Adams). M1 presents adjustment-stabilization, M2 and M3 present post-stabilization by $E$ and $F$, respectively

| Step-size | Method | $|\Delta E/E|$ | $|\Delta F/F|$ | CPU time (s) |
|---|---|---|---|---|
| 0.1 | Adams | 1.7E−3 | 2.8E−3 | 2.1 |
| | M2 | 1.1E−12 | 2.3E−5 | 2.5 |
| | M3 | 3.3E−4 | 1.1E−15 | 2.4 |
| | M1 | 2.2E−12 | 1.0E−15 | 2.5 |
| 0.05 | Adams | 5.5E−5 | 8.9E−5 | 4.1 |
| | M2 | 1.2E−15 | 8.7E−7 | 4.8 |
| | M3 | 1.0E−5 | 0 | 4.8 |
| | M1 | 2.5E−15 | 2.8E−17 | 4.8 |
| 0.01 | Adams | 1.6E−8 | 2.5E−8 | 20.1 |
| | M2 | 0 | 5.9E−10 | 23.5 |
| | M3 | 3.4E−9 | 2.8E−17 | 23.6 |
| | M1 | 0 | 2.8E−17 | 23.9 |

In this paper we point out the *post-stabilization* method just is the simplest example of the *adjustment-stabilization*. If only choosing one integral and adopting the accuracy-weight-matrix is unit matrix, the *adjustment-stabilization* is back to *post-stabilization*.

We do not discuss the case of the $P$ is not diagonal matrix, and we think it is very difficult to quantify the correlation of each two numerical algorithms. So we strongly advise to adopt numerical methods are same and independently to make $P$ is a unit matrix. In our numerical experiments, the numerical algorithms are classical one-step method and multi-step method, We think for other numerical algorithms are true too.

We find the manifold-correction method is not good in $N$-bodies problem. Wu et al. [29] pointed out the total energy in planetary dynamics does have certain effectiveness, but cannot yet completely raise the numerical precision. Wu adopt the individual Kepler energy (not a constant, but know their accurate reference value at any time) to modify the errors. We research this difficult, find the total angular momentum also facing same problem. But we cannot use individual angular momentum because we do not know the accurate reference value of each body at any time. And I think about how to apply the adjustment-stabilization in the 3-body (or $N$-body) is another deeply interesting topic.

## References

[1] K. Feng, in: K. Feng (Ed.), Proc. 1984, Beijing, Sym. Diff. Geometry and Diff. Equation, Science Press, Beijing, 1985, p. 42.
[2] J. Wisdom, Astron. J. 87 (1982) 577.
[3] J.M. Sarz-Serna, BIT 28 (1988) 877.
[4] J. Wisdom, M. Holman, Astron. J. 102 (1991) 1528.
[5] J.E. Chambers, M.A. Murison, Astron. J. 119 (2000) 425.
[6] J. Laskar, P. Robutel, Celest. Mech. Dyn. Astron. 80 (2001) 39.
[7] S.M. Kkola, P. Palmer, Celest. Mech. Dyn. Astron. 77 (2000) 305.
[8] H. Yoshida, Phys. Lett. A 150 (1990) 262.
[9] M.J. Duncan, H.F. Lerison, M.H. Lee, Astron. J. 116 (1998) 2067.
[10] H. Beust, Astron. Astrophys. 400 (2003) 1129.
[11] X. Wu, T.Y. Huang, X.S. Wan, Chin. Astron. Astrophys. 27 (2003) 124.

[12] X. Wu, T.Y. Huang, H. Zhang, X.S. Wan, Astrophys. Space Sci. 283 (1) (2003) 53.
[13] X. Wu, PhD thesis, The Department of Astronomy of Nanjing University, China, 2003.
[14] F.Y. Liu, X. Wu, B.K. Lu, Chin. Astron. Astrophys. 30 (2006) 87.
[15] J.D. Brown, Phys. Rev. D 73 (2006) 024001.
[16] J.D. Lambert, I.A. Watson, J. Inst. Math. Appl. 18 (1976) 189.
[17] G.D. Quinlan, S. Tremaine, Astron. J. 100 (1990) 1694.
[18] P. Moore, Celest. Mech. 17 (1977) 281–297.
[19] J. Baumgarte, Celest. Mech. 5 (1972) 490.
[20] P.E. Nacozy, Astrophys. Space Sci. 14 (1971) 90.
[21] H. Chin, PhD thesis, Institute of Applied Mathematics, University of British Columbia, Canada, 1995.
[22] E. Hairer, C. Lubich, G. Wanner, Geometric Numerical Integration, Springer, Berlin, 1999.
[23] X. Wu, T.Y. Huang, Chin. Astron. Astrophys. 29 (2005) 81.
[24] T. Fukushima, Astron. J. 126 (2003) 1097.
[25] L. Liu, X. Liao, Celest. Mech. Dyn. Astron. 59 (1994) 221.
[26] M.A. Murison, Astron. J. 97 (1989) 1496.
[27] X. Wu, J.Z. He, Inter. J. Modern Phys. C, accepted for publication, 2006.
[28] X. Wu, J.F. Zhu, J.Z. He, H. Zhang, Comput. Phys. Comm. 175 (2006) 15.
[29] X. Wu, T.-Y. Huang, X.-S. Wan, H. Zhang, Astron. J. 133 (2007) 2643.